# Mono/Multi-material Characterization Using Hyperspectral Images and Multi-Block Non-Negative Matrix Factorization

Mahdiyeh Ghaffari[1]*, Gerjen H. Tinnevelt[1], Marcel C. P. van Eijk[2], Stanislav Podchezertsev[2], Geert J. Postma[1], Jeroen J. Jansen [1]*

*Abstract*—Plastic sorting is a very essential step in waste management, especially due to the presence of multilayer plastics. These mono/multi-material plastics are widely employed to enhance the functional properties of packaging, combining beneficial properties in thickness, mechanical strength, and heat tolerance. However, materials containing multiple polymer species need to be pre-treated before they can be recycled as mono-materials and therefore should not end up in mono-material streams. Industry4.0 has significantly improved materials sorting of plastic packaging in speed and accuracy compared to manual sorting, specifically through Near Infrared-Hyperspectral Imaging (NIR-HSI) that provides an automated, fast and accurate material characterization, without sample preparation. Identification of multi-materials with HSI however requires novel dedicated approaches for chemical pattern recognition. Non-negative Matrix Factorization, NMF, is widely used for the chemical resolution of hyperspectral images. Chemically relevant model constraints may make it specifically valuable to identify multilayer plastics through HSI. Specifically, Multi-Block Non-Negative Matrix Factorization (MB-NMF) with "correspondence among different chemical species" constraint may be used to evaluate presence or absence of particular polymer species. To translate the MB-NMF model into an evidence-based sorting decision, we extended the model with an F-test to distinguish between mono-material and multi-material objects. The benefits of our new approach, MB-NMF, were illustrated by the identification of several plastic waste objects.

## I. INTRODUCTION

A major challenge in plastics recycling is the sorting of mixed plastic waste into streams that can be further processed into molecularly homogeneous recycling [1-3]. Each piece of plastic waste needs to be chemically characterized in very little time. Specifically challenging are multi-material plastics, which make up a significant fraction of plastics waste. Multi-layered packaging material consists of at least one plastic layer, with one or more layers of other polymers or e.g. paper, aluminum foils, polymers, or other materials [1, 4, 5]. Multi-layered plastic thereby contains plastics and other thin laminated material layers that require additional physical separation into constituent layers [6, 7]. Multilayer plastics are widely used in the food industry, as they may increase shelf-life [1-3]. Therefore, their specific recycling treatment requires accurate characterization [8-10] during sorting, which requires advanced, fast and dedicated data analyses that do not yet exist. Near InfraRed-HyperSpectral Imaging [11] is the most widely used analytical paradigm for fast industrial characterizations in automated plastics sorting. Chemometrics [12-14] plays a key role in the translation of spectroscopic data into chemically supported sorting decisions [9, 15-19]. Van Den Broek, W., et al., have used a remote sensing spectroscopic near-infrared (NIR) system for real-time plastic identification in mixed household waste [15]. First, the measurement setup was used for the acquisition of the spectroscopic image data and then a non-linear transformation is performed by a neural network for the supervised classification of these measured images[15]. Other authors have developed further dedicated chemometrics pipelines for plastics sorting. None of these methods however, enable characterization of multilayer materials. Mid-infrared hyperspectral images of the multilayer polymer film (MLPF) cross sections are acquired with a high-speed quantum cascade laser (QCL) based mid-infrared microscope and analyzed using different data analysis techniques. The investigated MLPF contains polypropylene and ethylene-vinyl alcohol co-polymer composite which is commonly used for food packaging. Principal component analysis (PCA) and multivariate curve resolution- Alternating Least Square (MCR-ALS) algorithms were used by Zimmerleiter, R., et al. to extract information about the object [17]. However, cross-section characterization of multilayer plastics is impractical and difficult to implement in online plastic sorting.

In another work, HSI-based sorting is investigated by Bonifazi G., et al. to perform an automatic separation of paper, cardboard, plastics, and multilayer packaging. The built Partial Least Squares – Discriminant Analysis model was able to characterize multilayer multi-material objects [22]. However, a discriminant analysis model requires a training dataset with all types of multilayer plastics with different compositions. In the case of new polymers, which are not included in the training database, the model will wrongly classify the object as belonging to a polymer class present in the training database. This highlights one of the cornerstones in the current paradigm

[1]*Radboud University, Institute for Molecules and Materials, Analytical Chemistry, P.O. Box 9010, 6500 GL Nijmegen, the Netherlands*

[2]*National Test Centre Circular Plastics (NTCP), Duitslanddreef 7, 8447 SE Heerenveen, the Netherlands*

of automated plastics sorting: it is treated in these existing approaches as a sorting task among a closed set of polymer species. Although contemporary plastics packaging developments continuously introduce new multilayer materials and even new polymer species. Non-negative matrix factorization [20, 21] of HSI data transcends the closed classification task as it provides a free estimation of compound spectra that can be supervised by (partial) spectroscopic, chemical and materials knowledge on known and unknown polymer materials. The method is based on multiplicative updates with the inherent non-negativity property, introduced by Lee and Seung [21] in the factorization of facial images. A known issue in NMF is the non-uniqueness of the obtained models when knowledge introduced into the model as model constraints is incomplete, limiting the correspondence between model results and actual chemistry. To increase possible uniqueness and resulting correspondence to actual chemistry, additional constraints may be implemented in NMF as "Correspondence among the Species" [22-24]. This correspondence constraint requires multiple single data sets that share a common mode and therefore multi-block data sets adds more information about the minor compounds in some blocks on one hand and helps to break the possible rank deficiency or highly collinear contribution maps of plastics on the other hand. This constraint imposes information about the presence/absence of components by defining the number and identity of compounds in each data block along the multi-block data. This additional information may strongly improve the accuracy of results. Furthermore, this data structure helps to generate a proper initial estimate for further decompositions.

The Multi-Block Non-Negative Matrix Factorization under the "correspondence among the species" constraint we propose here may be a highly suitable digital solution to translate HSI images of plastic waste into sorting decisions on multilayer plastics. This constraint is essential to tackle the high collinearity among NIR spectra of several widely used polymer species. We also introduce an F-test on the MB-NMF model observations to obtain crisp evidence-based sorting decisions for each plastic waste object into mono or multi-material streams. MB-NMF can accurately characterize objects (partly) composed of unknown polymer materials, which opens up the materials classification into an observation of new polymer and other materials occurring in plastic waste streams. We demonstrate this new digital method with numerical simulations and several real HSI data sets of real plastic waste objects with known polymer composition.

## II. MATERIALS AND METHODS

**A brief description of Non-Negative Matrix Factorization:** Non-negative matrix factorization (NMF) decomposes a matrix (which can be an unfolded HSI image), $\mathbf{R}_{IJ \times K}$, into the product of smaller submatrices $\mathbf{W}_{IJ \times n}$ and $\mathbf{H}^T_{n \times K}$:

$$\mathbf{R}_{IJ \times K} = \mathbf{W}_{IJ \times n} \mathbf{H}_{n \times K} + \mathbf{E}_{IJ \times K} \tag{1}$$

where $IJ$, $K$, and $n$ are the number of pixels and wavelengths, and chemical components (factors), respectively. To this, formalize NMF optimizes the following cost function:

$$\text{Minimize } (\mathbf{W}, \mathbf{H}) \quad \|\mathbf{R} - \mathbf{WH}\|_F \tag{2}$$
$$\text{subject to} \quad \mathbf{W} \geq 0, \mathbf{H} \geq 0$$

The multiplicative update algorithm is more popular due to the simplicity of implementation while a vast majority of algorithms are introduced in the literature. The updating parts are:

$$\mathbf{H} \leftarrow \mathbf{H} \frac{(\mathbf{W}^T \mathbf{R})}{(\mathbf{W}^T \mathbf{W} \mathbf{H})} \tag{3}$$

$$\mathbf{W} \leftarrow \mathbf{W} \frac{(\mathbf{R} \mathbf{H}^T)}{(\mathbf{W} \mathbf{H} \mathbf{H}^T)} \tag{4}$$

Equations 3 and 4 denote element-wise multiplication. These update rules preserve the nonnegativity of $\mathbf{W}$ and $\mathbf{H}^T$, given that $\mathbf{R}_{IJ \times K}$ is element-wise non-negative.

The resulting $\mathbf{W}$ and $\mathbf{H}^T$ can be unique or not, depending on the data structure. Using additional constraints is more convincing to lead the unique solutions. To achieve chemical/physical uniqueness, it is necessary to use information that is based on the chemical/physical structure of the data sets; like trilinearity, zero region/ selectivity, and correlation constraints.

The "Correspondence among the species" constraint is an important constraint that may lead to accurate solutions in multi-block datasets that may be imposed upon $\mathbf{W}$ in MB-NMF because the presence/absence of the compounds is known in some of the data blocks in advance. In the absence of this constraint the solutions my lack the accuracy and lead to incorrect interpretations. This prior knowledge also makes that the number of components in the MB-NMF model is known in advance because to determine the number using Singular Value Decomposition is too time-consuming for HSI-based online sorting. In principle, the number of components is equal to the number of known data blocks augmented to the unknown data block plus one. The extra component will be used to deal with the objects with unknown composition.

**Multi-Block Non-Negative Matrix Factorization:** Although unfolded HSI images may be analyzed by various curve resolution methods to extract pure contribution maps of single polymer species, a lack of enough information/constraint may often lead to a lack of uniqueness between species and therefore wrong interpretations and characterizations. Additional multi-block constraints (Trilinearity, correspondence among the species, area correlation constraint). Fusing HSI data from *unknown* mono/multi-material plastic with hyperspectral images of mono/multilayer plastic objects of *known* polymer composition, produces multi-block/higher order data which can be analyzed by MB-NMF as visualized in Fig. 1. Correspondence among species may be used as a constraint on the *known* blocks to impose the presence/absence of polymer species in every considered block. Prior to analyzing the data by MB-NMF, specific data pretreatments are necessary to focus upon the linear spectroscopic response of the polymer-containing pixels. Fig. 1b summarizes all the necessary steps. Object detection is necessary to remove most of the profiles corresponding to the conveyor belt and other pixels that contain non-polymer noise. In this work we used "activecontour" function in MATLAB software (R2021a) to segment each image into the foreground and background, leading to a binary output matrix indicating object pixels, removing most of the conveyor belt and selecting object pixels. Then we removed Saturated Spectral profiles (SS) with zero slopes in the middle part of the spectra (1100 – 1500 nm). The next two steps transform raw reflectance data into positive absorbance data. The Minus logarithm of the spectral profiles is calculated and then Asymmetric Least Squares is used to remove the baseline. A final Singular Value Decomposition, SVD then removes most of the remaining non-polymer noise. First, the data analyzed by SVD and then is reconstructed by a few Principal Components (PC). These steps are summarized in Fig. 1a and visualized in Fig. 2.

The data can be divided into two parts, known blocks, and unknown blocks. The "*known*" blocks which contain the data sets of known objects with known composition which can contain one or more compounds and their composition is known in advance. "Unknown" blocks are coming from unknown objects. This data structure helps to apply correspondence among the species constraint. To impose the mentioned constraint in MB-NMF, it suffices to set the zero values in the initial estimates (**W**). The resulting **W**, contains characteristic information about the composition of the unknown object. The sum of squares of elements of each column of **W** (variance) in the last block (came from the unknown object) is a good representative of the signal contribution of each polymer in the unknown object. The signal contribution (variance) can be used to distinguish mono-material and multi-material objects based on a statistical F-test. The F-test is generally used in statistics to determine if the standard deviations of two populations are equal. In this case, in the first step the variance of all polymers in the last data block is calculated. Then each value was divided to the minimum value. The minimum value is a good representation of noise population. Finally, a value of $P < 0.05$ was considered statistically significant. Thus, in the case of monolayer materials, only one polymer type will pass the F-test but for multilayers, more than one.

However, one final step is necessary to visualize the composition of the object which is sorted as multilayer, is a multilayer object or multicomponent object (with connected parts) which is made of different types of polymers like a water bottle (the cap could be PP or PE and the body is PET). First of all, the predestined columns of **W** will convert to binary numbers and then reshape to generate the contribution maps of the objects. Based on these maps it will be possible to distinguish between the connected objects (like a water bottle) and the multi-materials. The last column/row of **W**/**H** is added to capture the concentration and spectral profile of any of the unmodeled compounds (brown in Fig. 1b).

**a)** 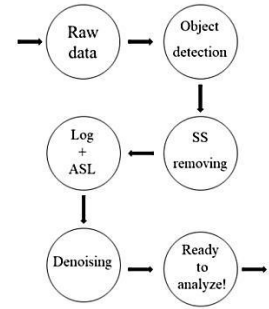

**b)** 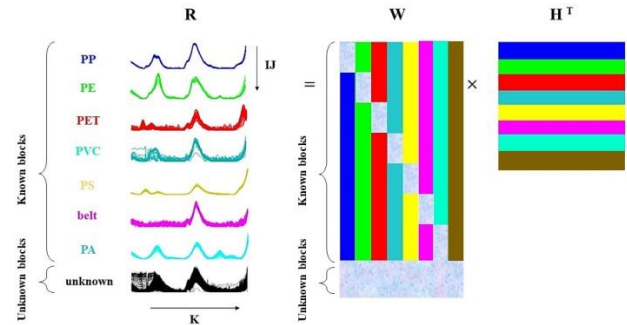

**Fig. 1.** Data preprocessing and analyzing are visualized in a and b, respectively. The data, **R**, can be decomposed to **W** and **H** using MB-NMF under correspondence among the species constraint. The known blocks are blue (PP), green (PE), red (PET), mint green (PVC), avocado green (PS), magenta (belt), and cyan (PA). The black color is used for the unknown data from an unknown object. Some of the unknown objects may partially contain new type of polymers. All information regarding this new material will collect as component X which is shown in brown.

Fig. 2 presents the effect of each data pretreatment step on the shape of the data for an example object. Fig. 2a is the recorded raw HSI of an optional object. Some spectral profiles should be removed from the data before data analysis of the signal-saturated spectral profiles. The corresponding spectral profile is horizontal and easy to remove by calculating the slope of all profiles in the middle of the data. Fig. 2b presents the normalized data after object detection and removal of some pixels (SS). The next step (c) is the data after baseline correction using asymmetric least squares on the minus logarithm of the data set. Finally, the data set is reconstructed by a few principal components for denoising purposes.

a)

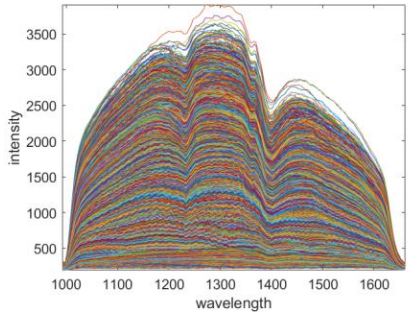

b)

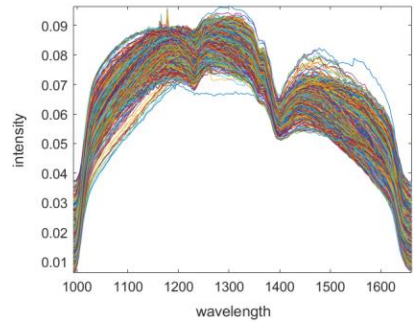

c)

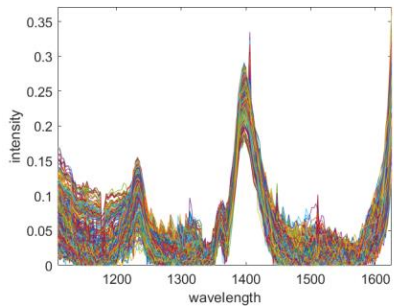

d)

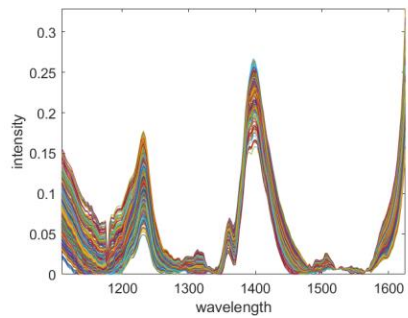

**Fig. 2.** The data set corresponds to the first unknown object that is visualized after some pretreatment. The raw data (a), after object detection and removing some pixels due to signal saturation (b), baseline correction of the minus logarithm of the data (c), and the data denoised by Singular Value Decomposition. The wavelengths are in nm.

### III. DATA DESCRIPTION

**Simulated data sets:** The objects which are used to make the known data blocks for the simulated cases are made of PET, PolyEthylene Terephthalate (first block), PP, PolyPropylene (second block), and PE, PolyEthylene (third block). These data blocks will be used as "known blocks" with known chemical composition (Fig. 3a). The mean image of the unknown objects which are multi-material plastics made of PP/PE (both known compounds) and PET/X/PP is presented in Fig. 3b, in which compound X is a new polymer that is not included in the known data blocks.

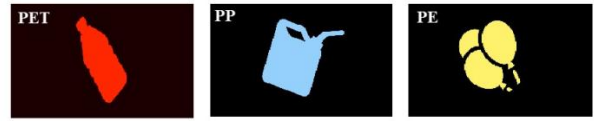

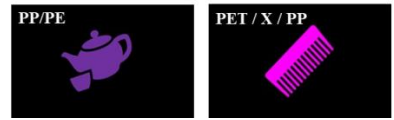

**Fig. 3.** a and b are the mean images of the generated HSI data for the known and unknown objects, respectively.

**Experimental data sets:** Some monolayer objects with known composition were collected as standards from the packaging waste. These objects are made of PP (Polypropylene), PE (Polyethylene), PET (Polyethylene terephthalate), PVC (Polyvinyl chloride), PS (Polystyrene), and PA (Polyamide). These objects will cover the most possible cases of multilayered plastics. For each object, HSI was recorded in NIR (900-1700nm). The recorded HSIs will be used to generate the known blocks of the data for MB-NMF. The mean spectral images are presented in Fig. S1. Furthermore, some objects were used as unknown objects. However, the characteristic information is provided by the brand owners in advance. These objects are made of PP/PE, PP/PS, PE, PP, PS, PVC, PET, PP/PE, PE/PET, respectively.

### IV. RESULT AND DISCUSSIONS

**Simulation:** The simulated data sets were analyzed by MB-NMF under the "correspondence among the species" constraint [22-24]. The signal contribution (variance) of each component in the last block were calculated and visualized in Fig. 4 for both data sets. The first unknown object is sorted as a multilayer/multi-material plastic made of PP and PE (Fig. 4a). In the second simulated data, the object contains PET, X, and PP (Fig. 3b). Compound X, a new type of polymer, is not represented in the known data blocks. By analyzing this multi-block data by MB-NMF, this object will be characterized as a multilayer/multi-material containing PET and PP. The presence of a new type of polymer that was not included in the model, does not wrongly identify the known polymer types in the object; it is sorted out as a multilayer/multi-material which contains PET/PP.

**a)**

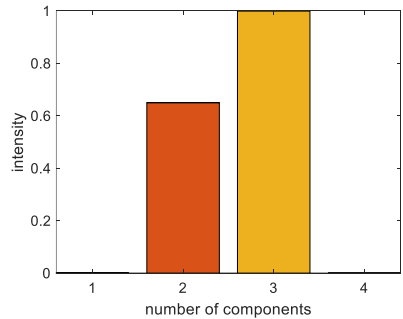

**b)**

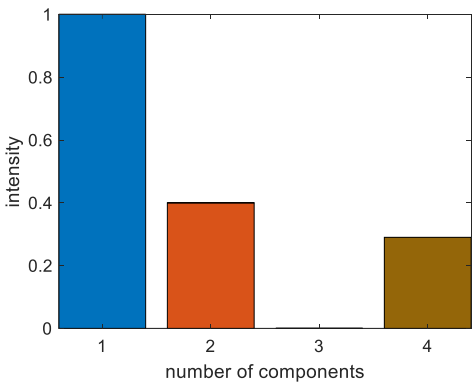

**Fig. 4.** The signal contribution is visualized versus the number of target polymers for the simulated data sets. Blue, red, yellow, and brown colors are used for PET, PP, PE, and X, respectively. In the bar plot, the intensity is an indication for the concentration of each polymer type. The numbers one to four is used for PET, PP, PE, and X, respectively.

**Experimental:** To analyze the experimental data sets, some additional data pretreatments were necessary compared to the simulated studies, as visualized in Fig. 1. The six different polymers and data from the empty conveyor belt (to impose the background) produce the known part of the multi-block data for MB-NMF. The selected Region Of Interest (ROI) is indicated in Fig. S1 by blue squares. The mean spectral profile of the ROIs, which is presented in Fig. S2, is used to generate the initial estimate for $\mathbf{H}^T$. The spectral profile of each polymer is shown by a specific color (Fig. S2).

The first object with unknown composition was used to visualize the effects of the used data pretreatments. The mean spectral image is presented in Fig. S3. removing the pixels corresponding to the belt (object detection) calculating a mask that removes pixels associated with the belt.

The multi-block data set corresponding to the first unknown object was analyzed by MB-NMF for characterization purposes in two different scenarios. In the first scenario, the known data blocks contain information about all possible polymers like PP, PE, PET, etc. The calculated variances for all of the polymers using the resulting $\mathbf{W}$ matrices are presented in Fig. 5. Based on this information, this object is a multilayer object which contains PE and PP. As it is clear in Fig. 5a, X does not have any variance which means there is not any unknown compound in this object.

In the other way (second scenario), the data set which contains PE is removed from the known blocks of the data to make an incomplete *known* block set. Based on the MB-NMF outputs, this object will be sorted as a multilayer, multi-material object made of PP and X. Thus the algorithm accurately predicts the *known* part of the object which is included in the known data blocks and is not disturbed by unknown interferents.

**a)**

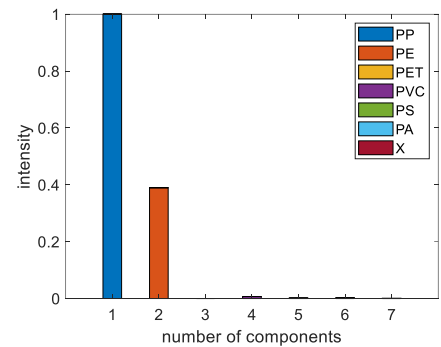

**b)**

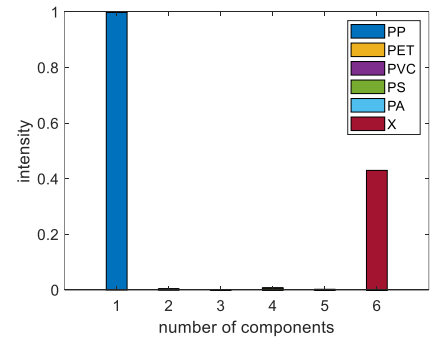

**Fig. 5.** The calculated signal contribution (variance) of all polymers for an unknown object by using MB-NMF under two different scenarios. With the aim of a statistical F-test with a 0.05 interval, these objects will be sorted as multi-material objects made of PP/PE (a) and PP/X (b).

The remaining HSIs of unknown objects were analyzed by MB-NMF after some data pretreatments as explained before. The generated bar plots using the last rows of $\mathbf{W}$ (corresponding to the last data block) are visualized in Fig. 6. The calculated variance (sum of squares of all elements) for each object is a way to characterize the mono/multi-material composition of all plastic objects. A statistical F-test with a 95 percent confidence interval was performed to evaluate the significance of the calculations. If more than one polymer type passes the F-test, the object is a multi-material plastic. Based on Fig. 6, a, g and h should be sorted as multi-material and the rest, as mono-material plastic. Finally, the recovered spectral profiles are visualized in Fig. 7.

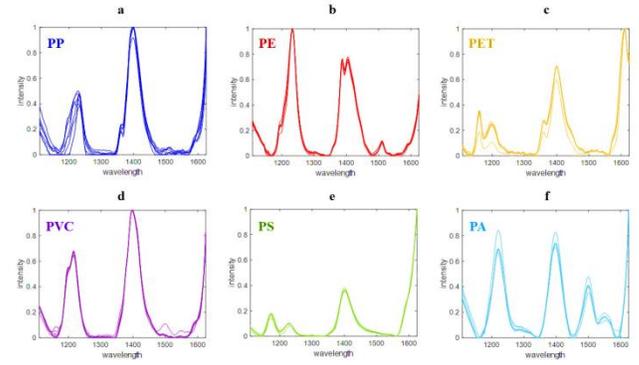

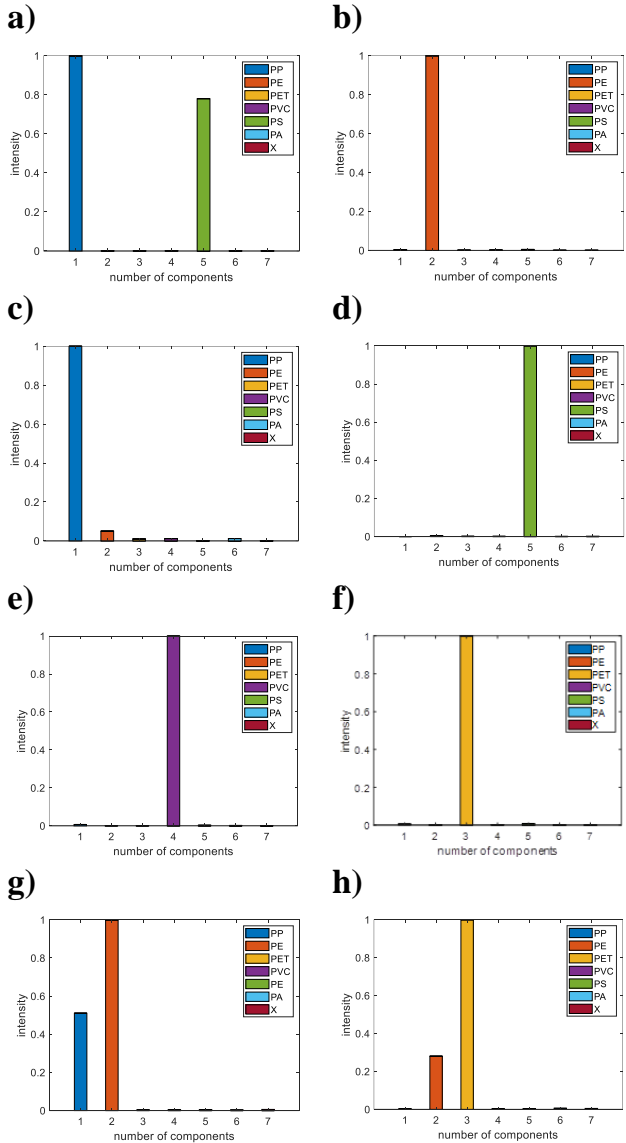

**Fig. 6.** The calculated signal contribution of all polymers in the unknown objects using MB-NMF.

**Fig. 7.** The resulting $\mathbf{H}^T$ Matrix using B-NMF for all of the unknown objects is visualized.

Based on the outputs of MB-NMF and the F-test, "a", "g" and "h" will be sorted as multi-material objects containing two different types of polymers. However, it is unknown yet whether these objects are made of two different polymers which are connected (like a water bottle whose body is PET and cap is PE/PP) or whether they are multilayer objects.

Thus, the contribution maps of $\mathbf{W}$ (which is converted to binary numbers first) are visualized for "g" and "h" in Fig. 8. As it is clear, object "g" which is visualized in Fig. 8a, is a multilayer object. However, object "h" which is made of PE and PET is an object with two parts (body in PET and cap in PE). There are some pixels in which PET and PE are on top of each other which is shown in white color. The black part in both maps is for the background.

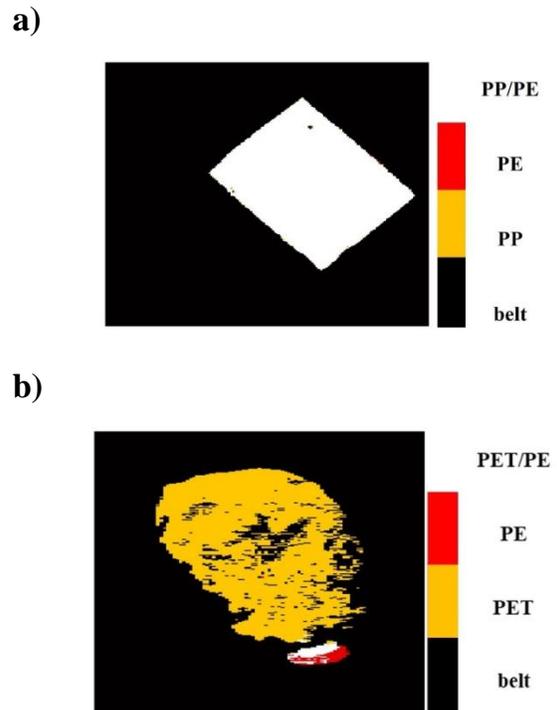

**Fig. 8.** This resulted in contribution maps of objects "g" and "h" using MB-NMF.

## V. Conclusion

Multilayer plastic characterization is demonstrated through the analysis of hyperspectral images using non-Negative Matrix Factorization. NMF can extract characteristic information from the Hyperspectral images using the non-negative property of the data. However, a lack of enough information/constraint can lead to wrong interpretations. Thus, additional information can be used to increase the accuracy of characterization and accurate interpretations consequently.

In this work, MB-NMF is used instead of NMF, because the multi-block structure enables the use of correspondence constraints which lead to uniqueness. Several known and unknown data blocks were fused in the first step. Then the fused data (MB) was analyzed by MB-NMF under correspondence among the different chemical species constraints. Besides accurate characterization, it will also tackle possible collinearity among concentration contribution maps. Another challenge of using data unmixing is the determination of the number of components. Analyzing the fused data instead of single data will tackle this issue. To use this algorithm in online plastic sorting the number of components is always equal to the number of known data blocks. MB-NMF can characterize the multi-material objects which partially is made of a new material/polymer that is not included in the known data block. In this case, the known part of the object will be predicted accurately due to unmixing. Thus this object will be sorted accurately.

## Acknowledgment

This project is co-funded by TKI-E&I with the supplementary grant 'TKI- Toeslag' for Top consortia for Knowledge and Innovation (TKI's) of the Ministry of Economic Affairs and Climate Policy. We thank all partners in the project "Towards improved circularity of polyolefin-based packaging", managed by ISPT and DPI in the Netherlands. It was partly funded by the Perfect Sorting Consortium, a consortium that develops AI technology to enable the intended use of sorting for the recycling of packaging material. The members are NTCP, Danone, Colgate-Palmolive, Ferrero, LVMH, Mars, Michelin, Nestlé, Procter & Gamble, PepsiCo, Ghent University, and Radboud University.

## References

1. Jönkkäri, I., et al., *Compounding and characterization of recycled multilayer plastic films.* Journal of Applied Polymer Science, 2020. **137**(37): p. 49101.
2. Lahtela, V., S. Silwal, and T. Kärki, *Re-Processing of Multilayer Plastic Materials as a Part of the Recycling Process: The Features of Processed Multilayer Materials.* Polymers, 2020. **12**(11): p. 2517.
3. Ruj, B., et al., *Sorting of plastic waste for effective recycling.* Int. J. Appl. Sci. Eng. Res, 2015. **4**(4): p. 564-571.
4. Wagner Jr, J.R., *Multilayer flexible packaging*. 2016: William Andrew.
5. Anukiruthika, T., et al., *Multilayer packaging: Advances in preparation techniques and emerging food applications.* Comprehensive Reviews in Food Science and Food Safety, 2020. **19**(3): p. 1156-1186.
6. Chen, X., et al., *Determining the composition of post-consumer flexible multilayer plastic packaging with near-infrared spectroscopy.* Waste Management, 2021. **123**: p. 33-41.
7. de Mello Soares, C.T., et al., *Recycling of multi-material multilayer plastic packaging: Current trends and future scenarios.* Resources, Conservation and Recycling, 2022. **176**: p. 105905.
8. Kaiser, K., M. Schmid, and M. Schlummer, *Recycling of polymer-based multilayer packaging: A review.* Recycling, 2017. **3**(1): p. 1.
9. Peñalver, R., et al., *Authentication of recycled plastic content in water bottles using volatile fingerprint and chemometrics.* Chemosphere, 2022. **297**: p. 134156.
10. Shen, L. and E. Worrell, *Plastic recycling*, in *Handbook of recycling*. 2014, Elsevier. p. 179-190.
11. Chang, C.-I., *Hyperspectral imaging: techniques for spectral detection and classification*. Vol. 1. 2003: Springer Science & Business Media.
12. Lavine, B. and J. Workman, *Chemometrics.* Analytical chemistry, 2008. **80**(12): p. 4519-4531.
13. Sharaf, M.A., D.L. Illman, and B.R. Kowalski, *Chemometrics*. Vol. 117. 1986: John Wiley & Sons.
14. Brown, S.D., et al., *Chemometrics.* Analytical chemistry, 1994. **66**(12): p. 315-359.
15. Van Den Broek, W., et al., *Plastic material identification with spectroscopic near infrared imaging and artificial neural networks.* Analytica Chimica Acta, 1998. **361**(1-2): p. 161-176.
16. Neo, E.R.K., et al., *A review on chemometric techniques with infrared, Raman and laser-induced breakdown spectroscopy for sorting plastic waste in the recycling industry.* Resources, Conservation and Recycling, 2022. **180**: p. 106217.
17. Zimmerleiter, R., et al., *QCL-based mid-infrared hyperspectral imaging of multilayer polymer oxygen barrier-films.* Polymer Testing, 2021. **98**: p. 107190.
18. da Silva, D.J. and H. Wiebeck, *ATR-FTIR Spectroscopy Combined with Chemometric Methods for the Classification of Polyethylene Residues Containing Different Contaminants.* Journal of Polymers and the Environment, 2022. **30**(7): p. 3031-3044.
19. Van den Broek, W., et al., *Application of a spectroscopic infrared focal plane array sensor for on-line identification of plastic waste.* Applied spectroscopy, 1997. **51**(6): p. 856-865.
20. Gillis, N., *Nonnegative matrix factorization*. 2020: SIAM.
21. Lee, D.D. and H.S. Seung, *Learning the parts of objects by non-negative matrix factorization.* Nature, 1999. **401**(6755): p. 788-791.
22. Ghaffari, M. and H. Abdollahi, *Duality based interpretation of uniqueness in the trilinear decompositions.* Chemometrics and Intelligent Laboratory Systems, 2018. **177**: p. 17-25.
23. Ghaffari, M., A.C. Olivieri, and H. Abdollahi, *Strategy to obtain accurate analytical solutions in second-order multivariate calibration with curve resolution methods.* Analytical chemistry, 2018. **90**(16): p. 9725-9733.
24. Ghaffari, M. and H. Abdollahi, *A conceptual view to the area correlation constraint in multivariate curve resolution.* Chemometrics and Intelligent Laboratory Systems, 2019. **189**: p. 121-129.